\documentclass{article}

\usepackage{arxiv}

\usepackage[utf8]{inputenc} % allow utf-8 input
\usepackage[T1]{fontenc}    % use 8-bit T1 fonts
\usepackage{hyperref}       % hyperlinks
\usepackage{url}            % simple URL typesetting
\usepackage{booktabs}       % professional-quality tables
\usepackage{amsfonts}       % blackboard math symbols
\usepackage{nicefrac}       % compact symbols for 1/2, etc.
\usepackage{microtype}      % microtypography
\usepackage{lipsum}
\usepackage{cite}
\usepackage[pdftex]{graphicx} 
\usepackage{amsmath}
\usepackage{graphicx}
\usepackage[section]{placeins}

\title{Defense in Depth: The Basics of Blockade and Delay}

\author{
  Andrew J. Lohn\\
  RAND Corporation\\
  Santa Monica, CA, 90407 \\
  \texttt{alohn@rand.org} \\
}

\begin{document}
\maketitle

\begin{abstract}
Given that individual defenses are rarely sufficient, defense-in-depth is nearly universal and options for individual defensive layers abound. We develop a simple mathematical theory that can help in selecting the type and quantity of defenses for two different defense-in-depth strategies: Blockade and Delay. This theoretical approach accounts for budgetary constraints and the number, skill, and diversity of attackers. We find that defenders have several reasons to be optimistic including that the number of required defenses increases more slowly than the number of attackers, that similar attackers are defended more easily than similar defenses are defeated, and that defenders do not necessarily need to act as quickly as attackers.
\end{abstract}

% keywords can be removed

\section{Introduction}
There is no perfect cyber defense, so providing high levels of security and surety is achieved by layering many separate imperfect defenses \cite{kuipers2006,nistSP800-57,stytz2004considering,shamim} in what is known as defense-in-depth. Of course, a layered set of imperfect defenses is also an imperfect defense, but a more expensive one, so some thought is justified in trying to understand how many defenses of what strength are best suited for the application at hand. Quantitative models exist for all sorts of defenses, \cite{mcqueen, luoAnalysis, wagner} attacks, \cite{santos, bilge} and vulnerabilities \cite{lohn, muegge} based on economics, \cite{kshetri, jentzsch, algarni} game theory, \cite{maleki, fielder, wang} or technical aspects of the systems, and even for modeling security budgets \cite{gordon, bohme} themselves. These models are most useful to information security officers and policy makers when they are at high levels of abstraction to understand how to select a mix of defenses that would constitute a defense in depth. While much previous work exists at high levels of abstraction, such as by considering cost \cite{guan, yin} to attacker and defender, we are not aware of any simple treatment of defense in depth. The elements of defense in depth are encapsulated in frameworks such as that provided by NIST \cite{nistFramework} but without a detailed model to illustrate tradeoffs and the interplay between the options and variables as we attempt to provide here. 

In this paper we discuss two defense-in-depth strategies depicted in Figure 1: Blockade and Delay. In a Blockade strategy the defender is trying to implement enough defenses that none of the individual attackers are able to to defeat or bypass all of the individual defenses. In a Delay strategy the attacker is meant to be slowed down enough to be discovered and expelled prior to achieving their malicious goal. 

\begin{figure}
  \centering
  \includegraphics[width=0.7\linewidth]{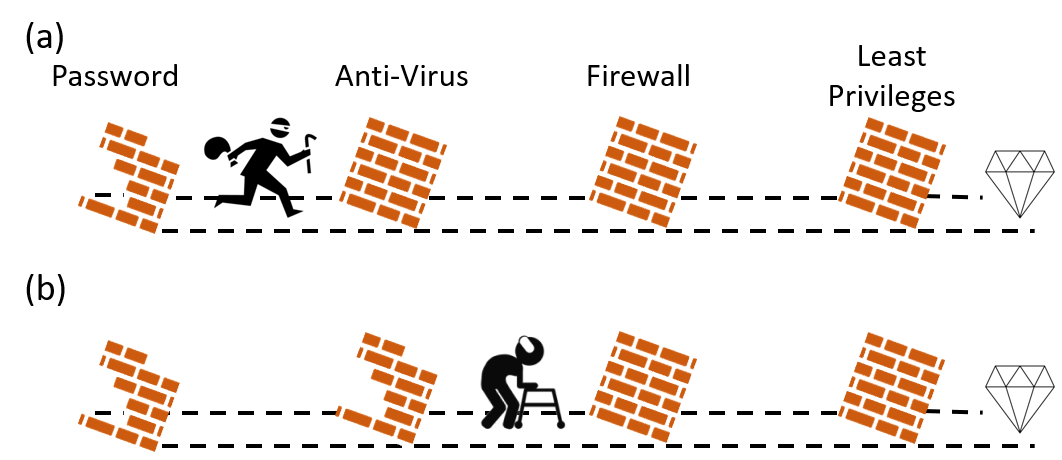}
  \caption{Two defense-in-depth strategies are illustrated. A Blockade strategy is shown in part a and a Delay strategy in part b.}
  \label{fig:fig1}
\end{figure}

The types of defenses and their relevant measures of performance are drastically different in the two cases. Defenses in a Blockade strategy simply need to have a low likelihood of being broken or bypassed. In a Delay strategy on the other hand, it is presumed that defenses will be broken or bypassed, and the measures of interest are the time it takes to do so, the detectability of an attacker during that process, and the ability for the defenses to be effectively reinstated (or potentially strengthened) after an attacker has previously defeated them. Examples of individual defenses in a digital security context include firewalls or authentications. Using password authentication as an example, in a Blockade strategy the likelihood that the password can be discovered, guessed, or bypassed would be the measure of interest, whereas for a delay strategy the measure of interest would be how long it takes to discover or bypass the password instead.

Once a specific defense-in-depth strategy has been selected, the appropriate number and types of defenses can be selected based on the anticipated threat (number, skill, and diversity of attackers) as well as the cost of the defenses. The rest of this paper walks through a simple foundational theory of both strategies. It is a quantitative approach, but it is not meant to provide exact specifications for defensive postures. Rather, it is meant to develop intuition concretely and clarify the interplay between the various factors and options.

\section{Blockade Strategies}
\label{sec:blockade}
\subsection{The Risk Equation}
If it is accepted that no defenses are perfectly impenetrable then risk minimization or acceptance becomes the focus in decisions about security posture. The risk of any event is simply the impact (I) of that event multiplied by the likelihood (L) of it occurring.

\begin{equation}
\label{eq1}
Risk = I * L
\end{equation}

Examples of impacts that might occur as a result of cybersecurity failures are brand damage \cite{whitler}, loss of intellectual property \cite{halbert}, or even physical damage to infrastructure \cite{ten}. These attacks are estimated to incur costs of hundreds of billion or trillions of dollars per year \cite{dreyer, csis, moar} but vary drastically from one organization to the next and the impacts are often outside the control of the organization. The likelihood of the impact being realized on the other hand is within their control through decisions about their defensive posture.

In a Blockade Strategy, an organization that is deciding on the appropriate defensive posture needs to consider that there are many possible attackers who are each trying to either get through or around many possible defenses. For any individual attacker to succeed, all of the defenses must fail. For the defender to succeed, all of the attackers must fail. 

Expressing that mathematically, suppose there are $N$ separate attackers, $n$ separate defenses, and each defense has a probability $p$ of being ineffective. For simplicity, we treat the defenses as all having the same probability of failure ($p$) and being statistically independent. These simplifying assumptions can be (and later will be) relaxed, but they are useful for developing simple relationships and intuition. For now, they lead to Eq \ref{eq2}a, where the probability of any individual attacker succeeding is the product of the probabilities of each defense failing.

\begin{subequations}
\label{eq2}
\begin{gather}
    P_{attacker succeeds}=\prod_{defenses}p=p^n \\
    P_{all attackers fail}=\prod_{threats}(1-P_{attacker succeeds})=(1-p^n)^N \\
    L = 1-P_{all fail} = 1 - (1-p^n)^N \\
    Risk = I[1-(1-p^n)^N]
\end{gather}
\end{subequations}

The probability that at least one of the attackers succeeds is simply one minus the probability that all attackers fail. Treating the different attackers as statistically independent as well, the probability that they all fail is the product of the probability of each of the individual attackers failing. Finally, this can be used to determine the likelihood of the damage occurring in the risk equation (Eq \ref{eq1}) , which is one minus the probability of all attackers failing.

\subsection{Likelihood of Breach}
\begin{figure}
  \centering
  \includegraphics[width=0.5\linewidth]{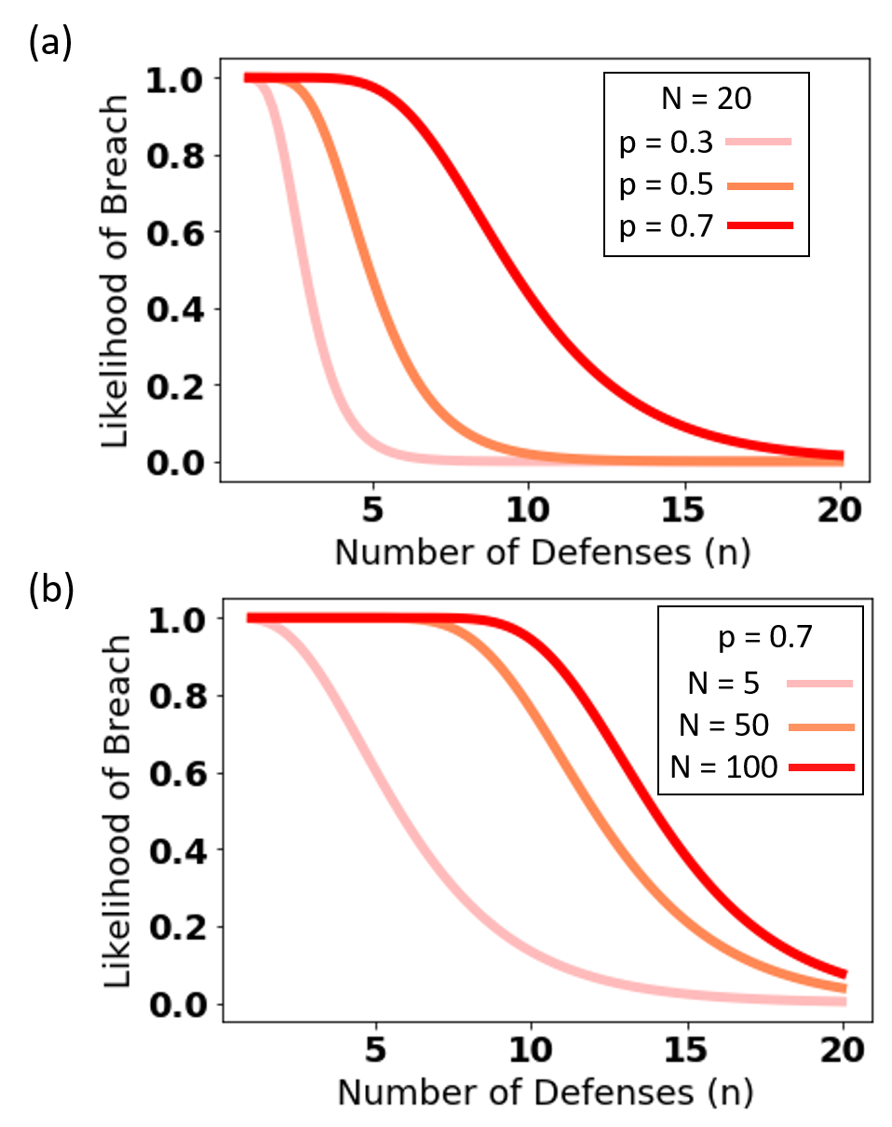}
  \caption{Likelihood of breach depends on number of defenses as well as (a) probability of those defenses failing and (b) the number of attackers.}
  \label{fig:fig2}
\end{figure}

There are differences in the types of concerns that various organizations face, some of which can be captured in these equations. An organization that faces many low-level attackers would have a large value for $N$ and a low value for $p$, whereas another organization may have fewer highly skilled adversaries. The same policies and postures might not be the right mix for both types of organizations. Militaries or critical infrastructure for example might only sustain a small number of attacks, but from highly skilled adversaries so they would be low $N$ and high $p$. Companies with valuable data on the other hand might encounter many relatively low-skill attacks and would be high $N$ and low $p$. Figure \ref{fig:fig2} illustrates the likelihood component of risk for various combinations of number of attackers, number of defenses, and probability of individual defenses failing.

As can be seen in Fig \ref{fig:fig2}a, to maintain the same likelihood of breach with weaker defenses (or stronger attackers) obviously means that a greater number of defenses are is needed. The number of additional defenses does not increase linearly though. The required number increases slowly at first but becomes faster with weaker and weaker defenses as can be seen from the wider gap between the red and orange curves than between the pink and orange in Fig \ref{fig:fig2}a. That is not the case when more attackers are added. More defenses are needed as more attackers are added in Fig \ref{fig:fig2}b, but as more and more attackers are added, the number of additional defenses required increases by less and less. This is a comforting indication that as the number of threats in the world continues to increase, it may be possible to maintain an effective defensive posture with only a relatively small number of additional defensive investments. 

\subsection{Optimizing Investments}
Since it is not possible to implement every security product being offered, then choosing the right mix of defenses becomes a balancing act between the number of defenses and their respective hardness. In that case, Eq \ref{eq2}c should be rearranged such that likelihood of breach becomes a parameter of the equation and we should solve instead for the number or hardness of the defenses as shown in Eq. \ref{eq3}. 

\begin{subequations}
\label{eq3}
\begin{gather}
    n=\log{1-p-(1-L)^{1/N}}\\
    p = [1-(1-L)^{1/N}]^{1/n}
\end{gather}
\end{subequations}

\begin{figure}
  \centering
  \includegraphics[width=0.5\linewidth]{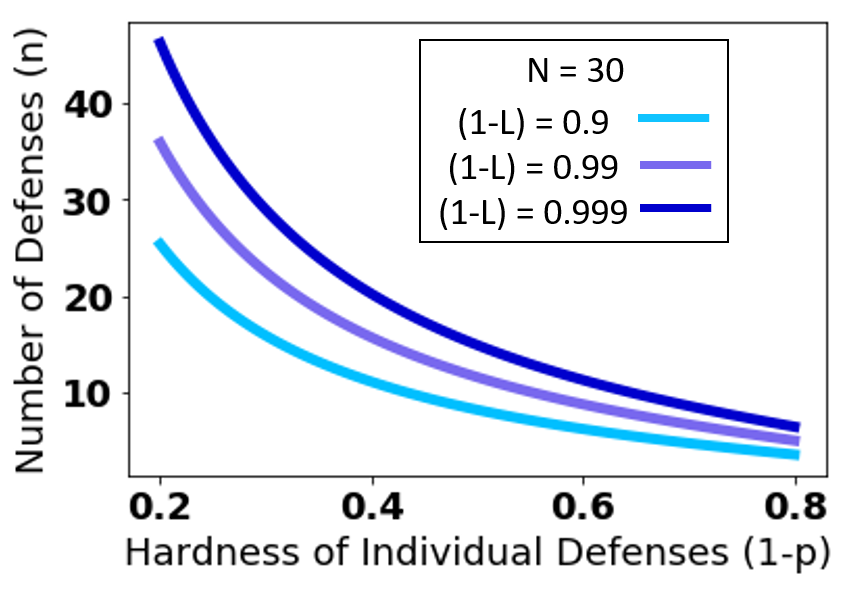}
  \caption{A given likelihood of breach can be achieved by many different defensive postures consisting of combinations of number and hardness of defenses.}
  \label{fig:fig3}
\end{figure}

Using either of these equations, it is possible to identify a number and hardness that provides the desired likelihood of breach. For a desired likelihood, there are many different ways to establish the defensive posture. One could use many weak defenses, few strong ones, or anything in between. Fig \ref{fig:fig3} shows the range of possibilities for achieving three different likelihoods. Borrowing a concept from economics, these are called indifference curves because any point along the curve will result in the same overall strength of defense and they illustrate the potential to trade between the hardness of defense and their quantity. If hardness and quantity were perfectly substitutable for each other then the curves would be linear, and they would be L-shaped if they were complementary such that adding more of one did not reduce the need for the other.

Of course, the defender may not actually be indifferent regarding the various points on the curves because they may not all have the same cost. To include that in the analysis though, one needs a concept of the cost of defensive postures. Those costs may not be known exactly but some aspects can be understood, and those traits can be used to create cost equations. The first trait is that weak defenses are expected to be inexpensive and strong defenses should be expensive.\footnote{There may be some defenses that are both effective and inexpensive, decisions to buy those do not require detailed economic analysis. To analyze further investments following the obvious investments, it is possible to separate out obvious ones. That is done by subtracting the cost of obvious defenses from the total cost and by dividing the acceptable likelihood of breach by the likelihood of breach using only those low-cost and effective defenses. The analysis then matches additional cost to additional likelihood of breach.}  It is also reasonable to think that no defense is perfectly impenetrable, so the best defenses should still have some non-zero probability of failure. 

\begin{figure}
  \centering
  \includegraphics[width=0.5\linewidth]{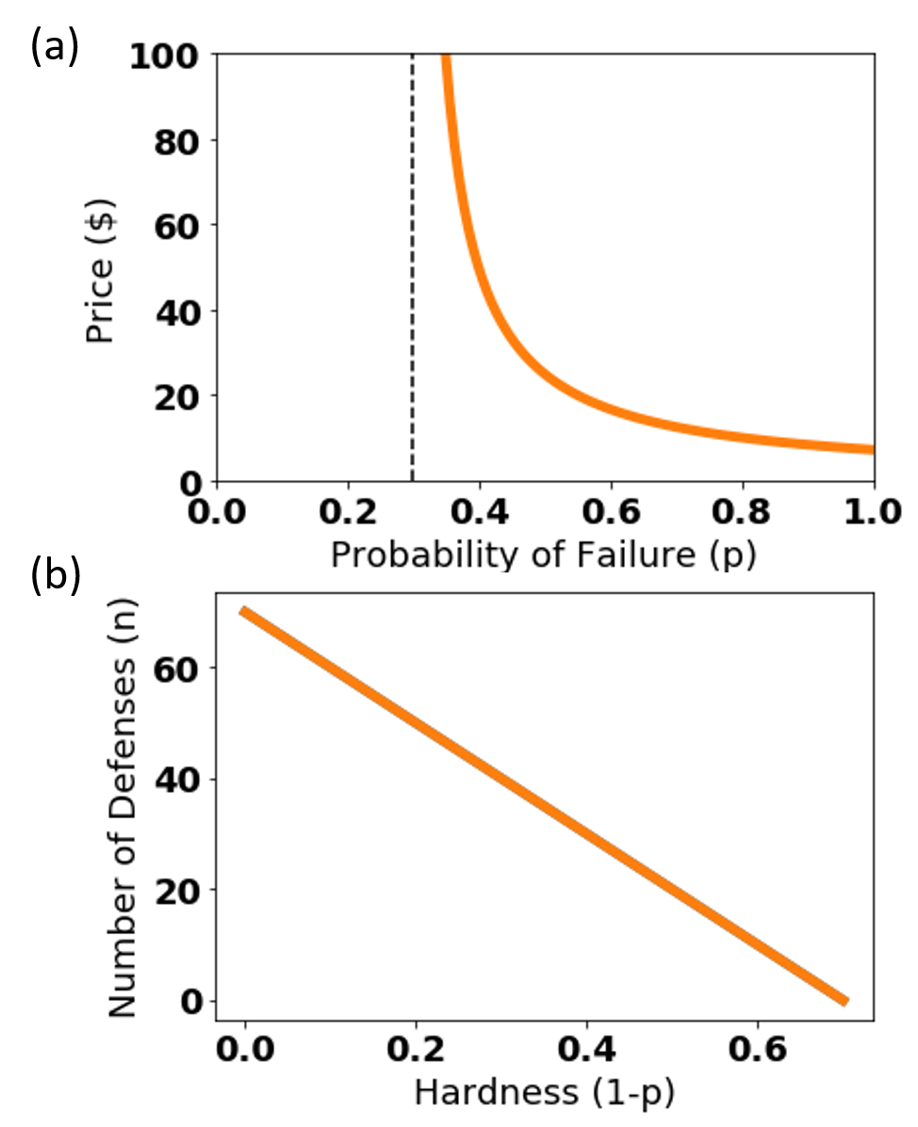}
  \caption{Price for individual defenses of varying efficacy (a) can be rearranged to give the range of various defensive postures that can be achieved the same cost.}
  \label{fig:fig4}
\end{figure}

A simple way to capture all these traits is shown in equation \ref{eq4} and Figure \ref{fig:fig4}a, where we let price be inversely related to probability of failure. The curve is shifted right by the probability that even the best defenses fail ($p_{best}$) so that no individual defenses are perfectly effective no matter how expensive they are.

\begin{subequations}
\label{eq4}
\begin{gather}
    P_{each} = \frac{A}{p-p_{best}} \\
    Cost = n * P_{each} \\
    n = \frac{Cost}{A}[p-p_{best}]
\end{gather}
\end{subequations}

The simple cost equation (Eq \ref{eq4}a) has $P_{each}$ as the cost for an individual defense, A as a scaling parameter that is the same for all defenses, and $p_{best}$ as the probability of failure for the best defenses. This equation can be rearranged (Eq. \ref{eq4}c) to match with the axes of the indifference curves. It is clear then that there are also many different possible defensive postures that can be envisioned, each having the same cost as shown in Fig \ref{fig:fig4}b. From a cost perspective, the defender is indifferent to changes along that curve, which, when plotted with respect to hardness of individual defenses becomes a straight line. That line is called the budget curve. Budget curves that correspond to higher costs are shifted up and to the right, whereas ones representing lower costs are shifted down and to the left.

\begin{figure}
  \centering
  \includegraphics[width=\linewidth]{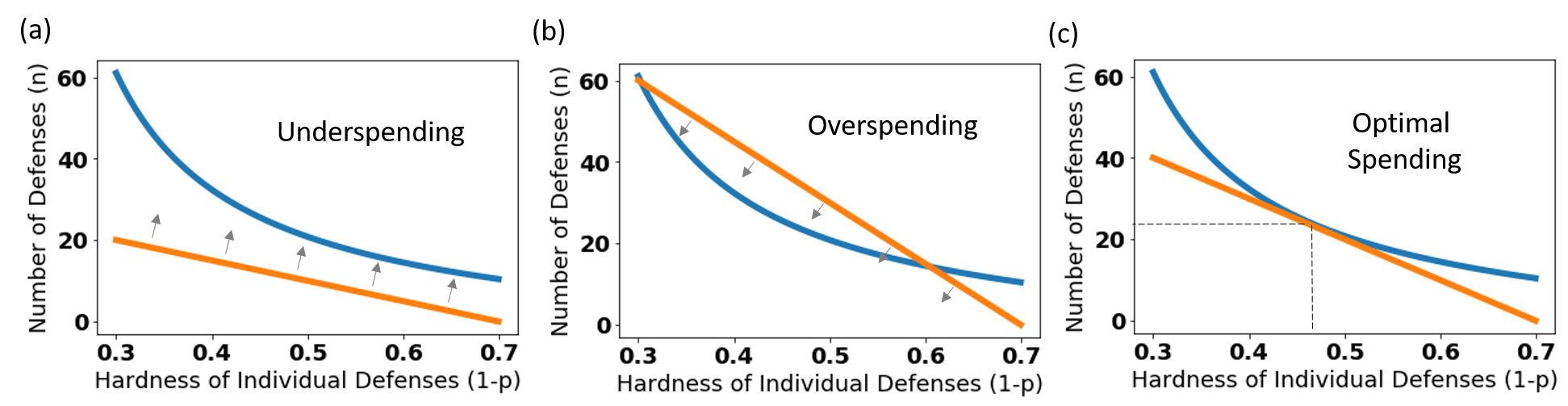}
  \caption{Indifference curves and budget curves can be combined to identify (a) underspending, (b) overspending, and (c) the optimal defensive posture.}
  \label{fig:fig5}
\end{figure}

With an indifference curve and a budget curve, it is possible to discuss optimal investment strategies that ensure the strongest set of defenses for a given budget or that minimize spending to achieve satisfactory levels of risk. Acceptable defensive postures exist wherever the budget curve and the indifference curve intersect. Fig \ref{fig:fig5}a shows a possible pair of indifference and budget curves that do not intersect. In this case, the budget is not large enough to meet the risk tolerance of the organization and more spending is needed to move the budget curve up and to the right. Fig \ref{fig:fig5}b shows a case where too much budget has been allocated. Acceptable defensive postures exist but other acceptable postures exist at lower cost. 

Fig \ref{fig:fig5}c shows the optimal spending and the corresponding number and hardness of defenses. In this example, the optimal defensive posture is one that contains a modest number of medium hardness defenses. Increasing either the number or hardness within the budget would result in defensive postures that lead to higher likelihood of breaches. Different budget curves would lead to different intersection points, which could be thought through similarly. It may not be possible to calculate precise budget curves or indifference curves for a real set of defenses but thinking through the process even roughly can help guide an assessment of whether the defensive budget is sufficient to meet the threat and whether it is best spent on a few strong defenses or a large number of weaker ones.

\subsection{Similarities Among Defenses and Attackers}
Thinking through the optimum defensive posture relies heavily on the indifference curves that were derived assuming statistical independence. We assumed that each of the defenses was not any stronger or weaker as a result of any of the other defenses and that none of the attackers were either. In reality, it is possible that some of the things that make one defense useful will also be present in the next defense, so the overall defensive posture might be weaker because the defenses are not actually independent. Addressing that assumption only requires a simple change to the derivation but first requires a brief discussion of statistical dependence. 

\begin{figure}
  \centering
  \includegraphics[width=0.5\linewidth]{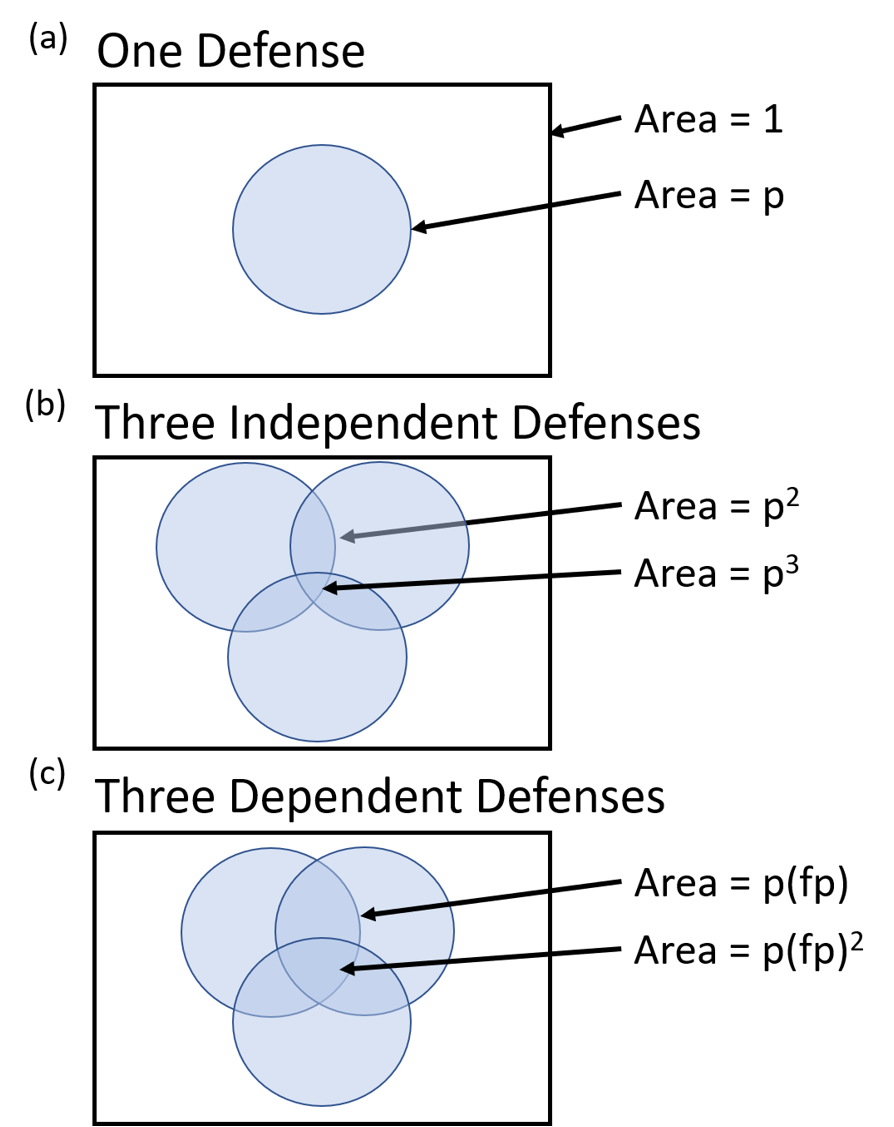}
  \caption{The probability of a single defense failing is shown in part a. Parts b and c show the probability of multiple defenses failing for statistically independent and dependent cases respectively.}
  \label{fig:fig6}
\end{figure}

Part a of the Venn diagram in Fig \ref{fig:fig6} shows only one defense, where the shaded fraction corresponds to the likelihood of that defense failing. There is more white space than shaded space so in this example it is more likely that the defense will succeed than fail. When there are multiple defenses, the quantity of interest is the probability that they all fail, which in a Venn diagram is depicted as the overlapping of shaded regions. For statistically independent defenses, the amount of overlap must equal the product of the probabilities of failure for each defense, as depicted in Fig \ref{fig:fig6}b. If the defenses are statistically dependent, though, the overlap can be larger than that (Fig \ref{fig:fig6}c). The larger overlapping region corresponds to a larger probability of the entire set of defenses failing. For this discussion we treat each defense as having constant overlap with one another to illustrate the concept and understand the fundamental behavior of dependence. In practice, some defenses will be more similar than others. The general cases can be easily calculated either numerically or by introducing distributions of dependence, but the basic intuition can be gained with just this simple approach.

Starting from the probability of one defense failing, the overlap for one additional defense can be found by multiplying by a dependence factor $f$, where $f$ is greater than one if the defenses are more likely to fail together and less than one if the existence of one defense makes the second less likely to fail. The dependence factor is multiplied again for every additional layer of defense as shown in Fig \ref{fig:fig6}c.

Introducing this dependence factor into the likelihood equation (Eq \ref{eq2}c), the $p^n$ term is multiplied by $f^{(n-1)}$, leading to Eq \ref{eq5}.

\begin{subequations}
\label{eq5}
\begin{gather}
    L = [1 - (1- f^{n-1}p^{n})^N] \\
    n = \log[f(1-p-(1-L)^{1/N}]
\end{gather}
\end{subequations}

\begin{figure}
  \centering
  \includegraphics[width=0.5\linewidth]{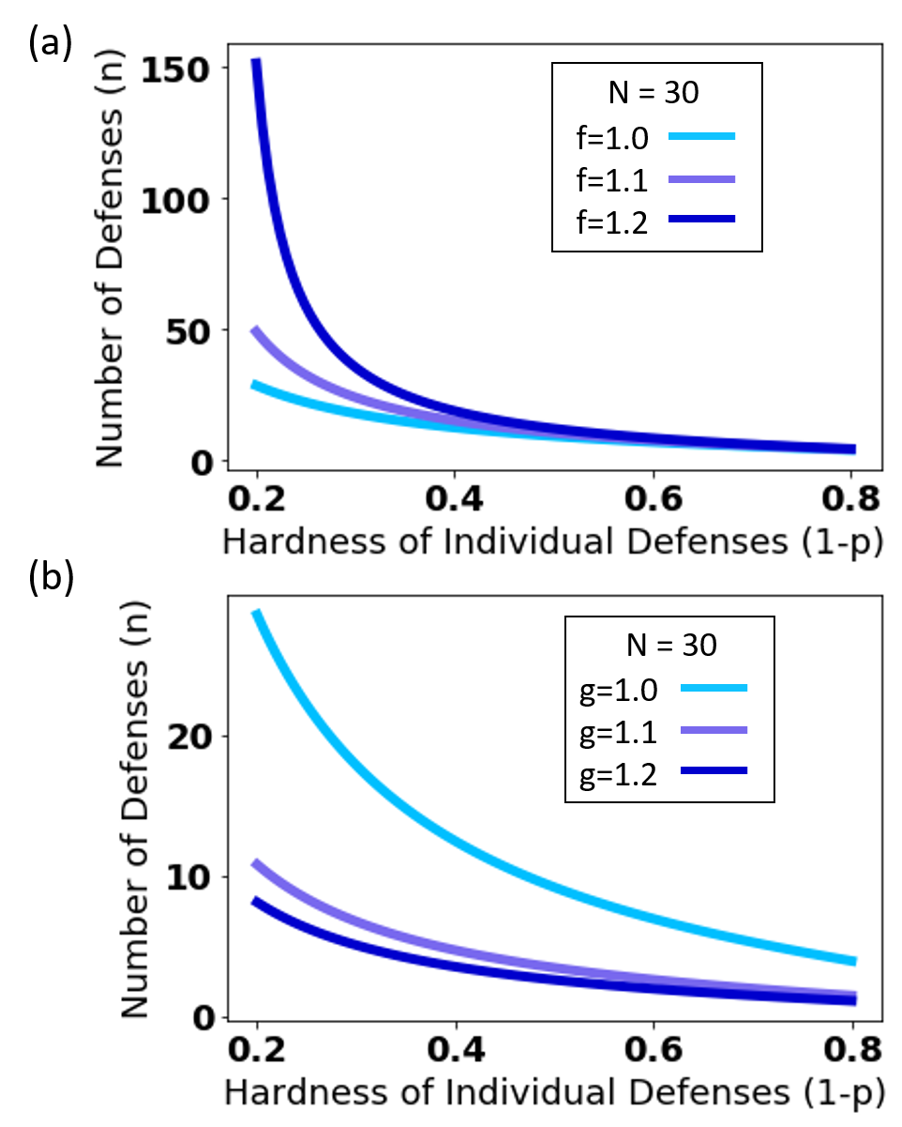}
  \caption{Statistical dependence in defenses has a large effect for many weak defenses and a strong offsetting effect among attackers.}
  \label{fig:fig7}
\end{figure}

Statistical dependence in defenses implies that more or better defenses will be needed to achieve the same level of risk. That can be seen in Fig \ref{fig:fig7}a which shows indifference curves shifting toward higher numbers and hardness to achieve the same likelihood when the dependence factor is increased. The shift is much more pronounced for weaker defenses than for hard defenses because weaker defenses require greater numbers of defenses to remain indifferent. The more defenses there are, the more times the dependence factor must be multiplied and therefore the more effect it has on the likelihood of breach. That makes the curve more L-shaped because quantity of defenses because a worse substitute for quality when their failures become more statistically dependent. 

In deriving the risk equation, we also assumed the attackers were statistically independent, but they can be dependent as well. The same process that was used for dependence in defenses can be followed for dependence in attackers. For the attacker dependence factor, we use the letter $g$ to differentiate from defense dependence $f$.

\begin{subequations}
\label{eq6}
\begin{gather}
    L= [1-g^{N-1}(1-f^{n-1}p^n)^N] \\
    n = \log[f(1-p-(\frac{1-L}{g^{N-1}})^{1/N}]
\end{gather}
\end{subequations}

Fig \ref{fig:fig7}b shows the indifference curve for statistically dependent attackers and statistically independent defenses. The curve shifts toward fewer and weaker defenses because each attacker is a little more likely to be foiled by the same defenses as the previous ones had. It is also apparent from the figure that statistical dependence among attackers leads to large reductions in the number and hardness of defenses required to achieve a desired likelihood of breach. The effect is much more pronounced for dependence between attackers than it is for dependence between defenses, which is another point of comfort tilting the battle in the defender’s favor. In practice, hackers form tight communities, and there is likely to be significant amounts of statistical dependence between attackers that can make defense substantially more achievable than would otherwise be anticipated.

\section{Delay Strategies}
\subsection{Time to Removal}
In a Delay Strategy, the battle may be better visualized as a back and forth process where the attacker is constantly trying to discover and defeat each defense, and the defender is constantly trying to discover and remove the attacker.  The defensive posture can be based on attack-defend timelines where four main processes are constantly taking place. The attacker is constantly probing or moving through the network and progressively defeating or bypassing defenses. For each defense, the attacker encounters two delays, one while the attacker is finding the next defense, and another while the attacker is breaking it. At the same time, the defender is constantly searching for broken or bypassed defenses and is actively trying to repair them, but both of those take time as well. 

In this simple scenario, one quantity of interest is the amount of time until there are no functioning defenses remaining, or alternatively, the number of functioning defenses at any given time. That can be found by knowing the number of defenses at the start and calculating the rate at which defenses are bypassed or broken.

\begin{equation}
\label{eq7}
    \frac{dn_B}{dt} = \frac{1}{\tau_F + \tau_B} - \frac{n_B}{\tau_D+\tau_R}
\end{equation}

In Eq \ref{eq7}, $n_B$ is the number of broken or bypassed defenses, $\tau_F$ and $\tau_B$ are the times required for the attacker to find and break defenses respectively, and $\tau_D$ and $\tau_R$ are the times required for the defender to detect and repair the broken defenses. The overall rate for broken defenses ($dn_B/dt$) is simply the rate at which defenses are broken minus the rate at which they are fixed. The second term is made proportional to $n_B$ because each broken defense provides another opportunity for the defender to discover a broken one. If there are no broken defenses then the rate at which they are discovered is zero, and if there are many broken ones then it becomes easier to find one. This presumes that each broken defense provides a continuing opportunity to discover the attacker which is true of some defensive postures and not of others, it is an assumption that we will follow for now and revisit shortly.

Integrating Eq \ref{eq7} provides the number of broken defenses at any given time.

\begin{equation}
\label{eq8}
    n_B = \frac{\tau_D+\tau_R}{\tau_R+\tau_B}[1-\frac{-t}{e^{\tau_D + \tau_R}}]
\end{equation}

\begin{figure}
  \centering
  \includegraphics[width=0.5\linewidth]{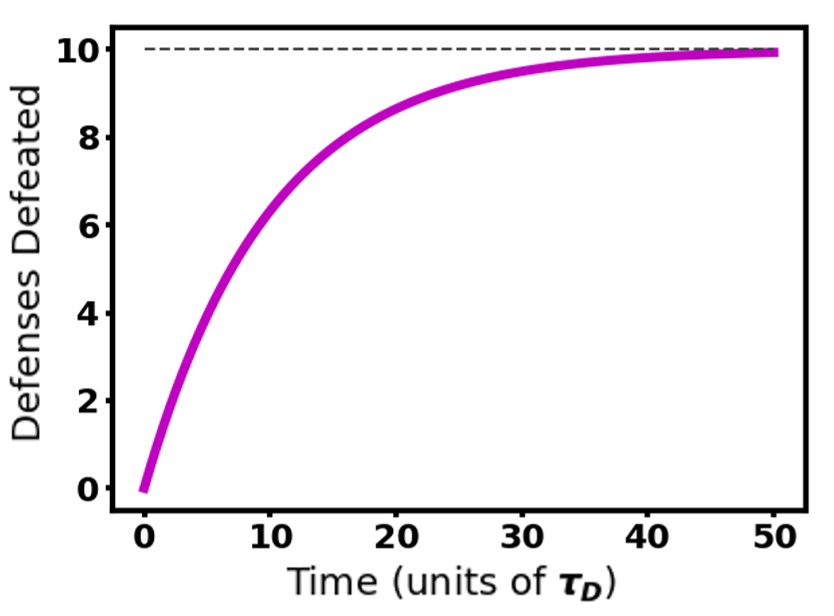}
  \caption{The number of defenses defeated grows with time but saturates at a finite value once the defenders can find broken defenses as quickly as attackers can create them.}
  \label{fig:fig8}
\end{figure}

Eq \ref{eq8} indicates that the number of broken defenses starts off at zero, then increases quickly at first but slows until it eventually reaches a maximum number. At that point, the defenders are able to discover and fix broken defenses as quickly as the attackers can break them. In Fig \ref{fig:fig8}, that maximum number of defeated defenses is 10.

This is another result that could come as a relief to defenders. It suggests that there is a number of defenses that is sufficient to defeat attackers and that none of those defenses need to be unbreakable, they just need to impose delays on the attackers. The defenders also do not have to act as quickly as the attackers, they just have to scale the number of defenses appropriately to account for the speed difference between the attacker and the defender.

\subsection{Number of Defense Required}
One could craft a defensive investment strategy by simply having more defenses than the maximum number calculated from Eq \ref{eq8} by setting $t=\infty$. Eq \ref{eq8} then reduces to Eq \ref{eq9}, which simply states that the number of defenses needs to be greater than the ratio of the time it takes a defender to discover and expel the attacker to the time it takes an attacker to break or bypass defenses. 

\begin{equation}
\label{eq9}
    n > \frac{time\ to\ discover}{time\ to\ break}
\end{equation}

This model provides some useful intuition and guidance, but the same simplicity that makes it intuitive also introduces caveats that should be considered. For example, this approach only addresses an individual attacker when there could be many. To adjust for more attackers, one could simply multiply the rate of broken defenses by the number of attackers ($N$) in Eqs \ref{eq8} and \ref{eq9}, which results in an increase in the number of defenses needed that is proportional to the number of attackers as shown in Eq \ref{eq10}.

\begin{subequations}
\label{eq10}
\begin{gather}
    n_B= \frac{N(\tau_D + \tau_R}{\tau_R + \tau_B} [1-\frac{-t}{e^{\tau_D + \tau_R}}]  \\
    n > \frac{N(time\ to\ fix)}{time\ to\ break}
\end{gather}
\end{subequations}

This scale of increase in number of defenses with number of attackers is less optimistic for defenders. It suggests that using large numbers of defenses and hoping to repair them faster than the attackers can defeat them - “playing whack-a-mole” – is untenable for organizations that face many threats unless the defenders can impose large time delays and maintain short discover and repair times. For organizations that only face few threats it may be a more satisfying option.

This time-based model was derived using another simplifying assumption that may not be reasonable in all cases. It assumes that all the previously broken defenses remain broken in a way that is discoverable by the defender. In some cases, it may be possible for the attacker to place footholds or backdoors that allow them to cover their tracks. In this case, the defender will only discover the attacker during the current stage of the attack and the previously broken defenses do not increase the detection rate.

\subsection{Attackers that Cover Their Tracks}
The probability of detecting the attacker while it is trying to defeat any individual defense depends on how long it takes for the defense to be defeated and how good the defender is at detecting attacks. Mathematically, that probability of detection at any individual defense can be found by specifying a probability density function for detection and integrating it over the duration of that stage of the attack as shown in Eq \ref{eq11}. A reasonable probability density function to consider is an exponential distribution which is the distribution that occurs when there is no reason to think the attacker would be any more or less detectable over time. 

\begin{equation}
\label{eq11}
    P_{Disc} = \int_{t=0}^{t=\tau_a}\lambda e^{-\lambda t}dt = 1 - e^{\lambda\tau_a}
\end{equation}

In Eq \ref{eq11}, $\tau_a$ is the time it takes an attacker to traverse to and defeat an individual defense, and $\lambda$ is the rate at which defenders can discover an attacker engaged in that process. That rate is simply the inverse of the average time to make those discoveries ($\tau_d$).

Similarly to earlier discussion of Blockade Strategies, the probability of an attacker not being discovered at any point is the product of them it not being discovered at any of the individual defenses.

\begin{equation}
\label{eq12}
    L = \prod_{Defenses} e^{-\lambda \tau_a} = e^{-\lambda \tau_a n}
\end{equation}

This likelihood has the same meaning as it did for the Blockade discussions, but in this case there are three variables that the defender may have some control over. They may make investments in tools or personnel to detect adversaries ($\lambda$), they can invest in defenses or technologies that slow the attacker ($\tau_a$), and they can choose to implement any number of those defenses ($n$). According to Eq \ref{eq12}, each of those options has exactly the same effect in terms of risk reduction because all three of the variables occur in the same place in the equation. The place where they occur together is also fortunate from a defensive standpoint. Because these variables are arguments of an exponential, improvements in any of the defensive variables result in exponential reductions in risk.

That is not to say that it is sufficient to focus on whichever risk reduction method is most convenient. The rate at which risk is reduced for an increase in one variable depends on the size of the other two variables as shown for changes in the number of defenses in Eq \ref{eq13}, which has both $\lambda$ and $\tau_a$ as prefactors.

\begin{equation}
\label{eq13}
    \dfrac{\partial(1-L)}{\partial n} = \lambda \tau_a e^{-\lambda \tau_a n}
\end{equation}

\begin{figure}
  \centering
  \includegraphics[width=0.5\linewidth]{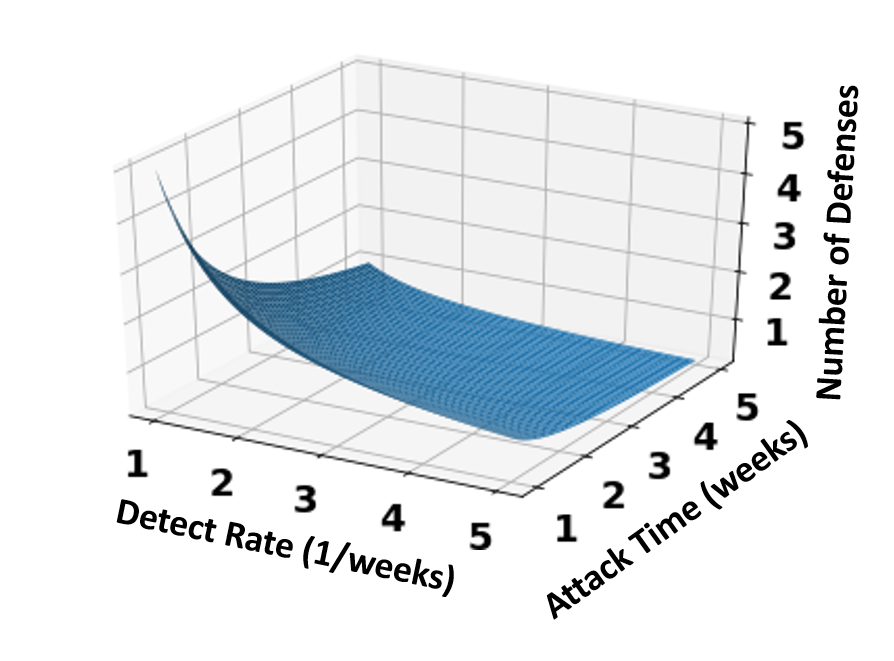}
  \caption{An indifference surface showing the possible combinations of detection rate, attack time, and number of defenses that lead to equivalent likelihood of breach.}
  \label{fig:fig9}
\end{figure}

Therefore, the number of defenses, detection rate, and attack duration should all be increased together to create a robust defensive posture. The exact values for those parameters will depend on risk tolerance and costs of the various defensive options just as was the case using the indifference and budget curves in the Blockade Strategy discussions. The difference for Delay Strategies is that, because there are three variables, the indifference curve is an indifference surface as illustrated in Fig \ref{fig:fig9}, and the budget curve would be a budget surface.

\begin{figure}
  \centering
  \includegraphics[width=0.5\linewidth]{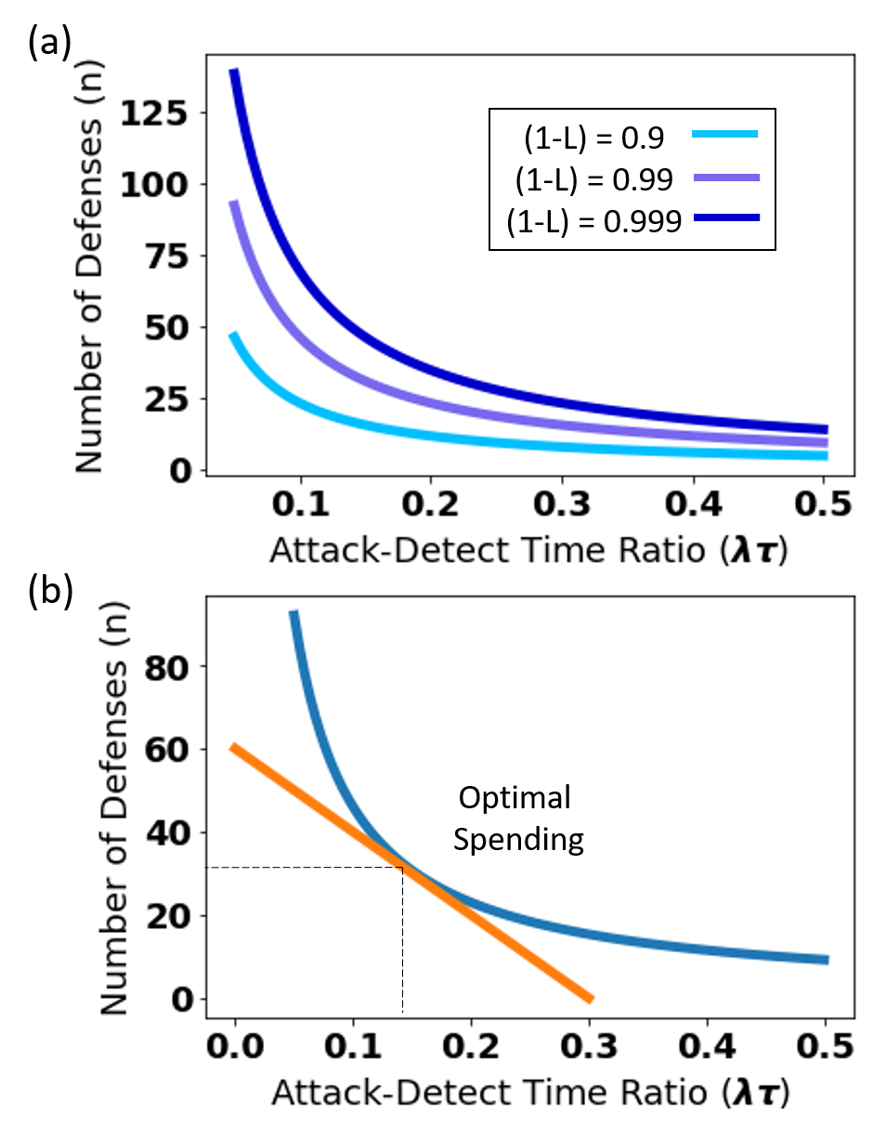}
  \caption{(a) Indifference curves and (b) optimal spending shown by compressing attack time and detect rate to a single variable.}
  \label{fig:fig10}
\end{figure}

Because three dimensional surfaces are difficult to visualize, the concept can be illustrated more compactly by compressing two of the variables together as shown in Fig \ref{fig:fig10}. Here the rate of detection and time for attacks are shown together on the x-axis. Small values of that combined $\lambda\tau_a$ variable indicate that detection is much slower than attacks.

The optimal defensive posture can then be found just as before and as illustrated in Fig \ref{fig:fig10}b while helping to avoid overspending and insufficient configurations or quantities of defenses.

\subsection{Similarities in Defenses and Attackers}
So far, this $L$ for Delay Strategies has only been considering a single attacker but there may be multiple attackers. Following the same logic as Eq \ref{eq2} from the start of the paper, though, this $L_1$ can be used to derive a likelihood for multiple attackers ($L_N$). This likelihood is one minus the probability that all attackers fail. And the probability that all attackers fail is the product of one minus the probability that a single attacker succeeds, which is the $L$ derived in Eq \ref{eq12}. This series of concepts is laid out in Eq \ref{eq14}a and then rearranged to create the new indifference curves for multiple statistically dependent attackers. 

\begin{subequations}
\label{eq14}
\begin{gather}
    L_N = 1- P_{all\ attacks\ fail} = 1-\prod_{threats}1-P_{attack\ succeeds} = 1- g^{N-1}(1-L_1)^N \\
    L_N = 1- g^{N-1}(1-e^{-\lambda \tau_a n})^N\\
    n =-\frac{\log[1-(\frac{1-L_N}{g^{N-1}})^N}{\lambda \tau_a}
\end{gather}
\end{subequations}

\begin{figure}
  \centering
  \includegraphics[width=0.5\linewidth]{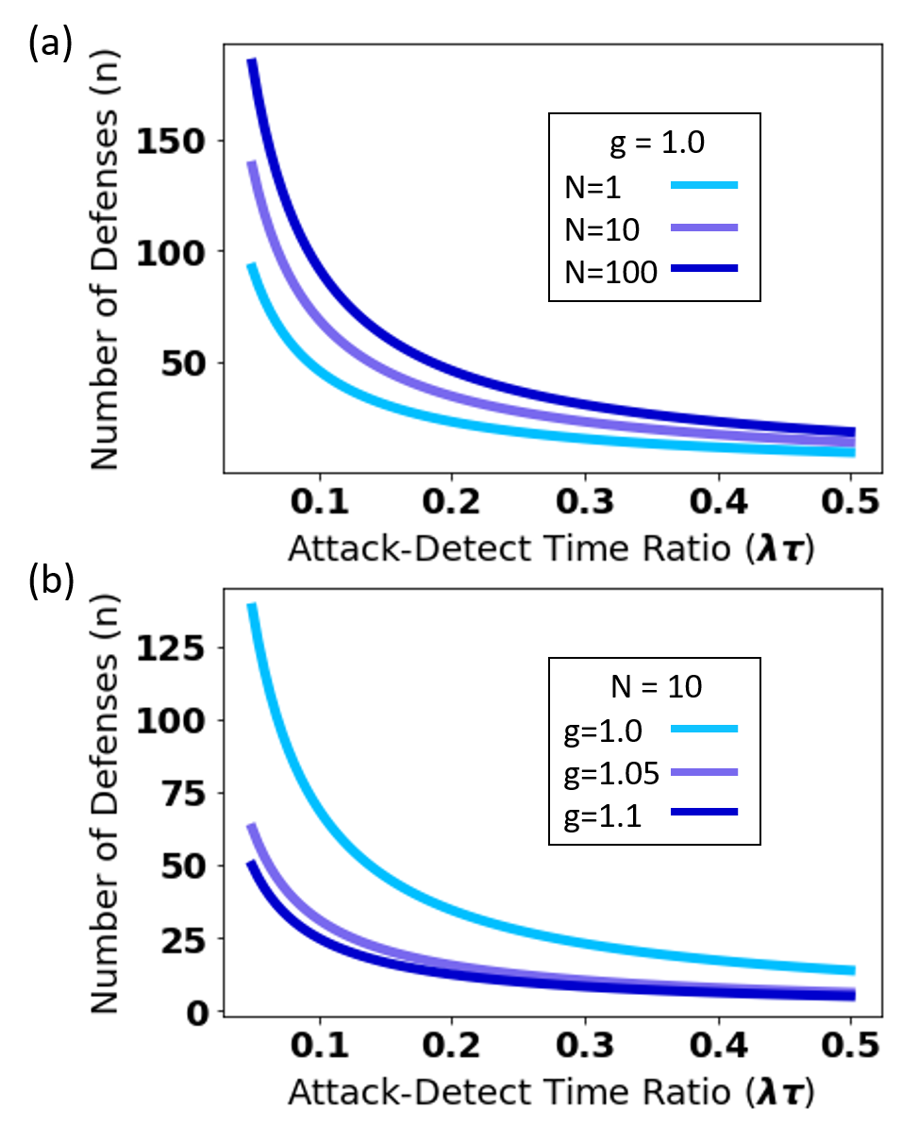}
  \caption{Compressed indifference curves showing slow increases in defensive posture required for growing number of attackers and substantial decreases for statistical dependence of those attackers.}
  \label{fig:fig11}
\end{figure}

Once again, increasing the number of attackers leads to only relatively slow increases in the required additional investments in defensive posture, in terms of number of defenses, rate of detection, and time for attacks. Statistical dependence among attackers again provides a larger boost to defenders, allowing for substantially fewer or weaker defense to achieve the same levels of risk. The assumptions leading to this final equation are fairly conservative, including multiple attackers and assuming that attackers are able to perfectly cover their tracks at every stage of the attack. Still, defensive postures that provide reasonable levels of assurance appear possible without requiring an impossibly large number or effectiveness of individual defenses. Again, although calculating precise curves may be challenging or impossible, thinking through these relationships can help guide the suite of defenses to match budgetary constraints for an anticipated threat model.

\section{Conclusions}
This paper has laid out a set of conditions and assumptions describing two different defense-in-depth strategies. By mathematically specifying these conditions, which are individually simple but together more complicated, it has been possible to understand how variables such as number and hardness of defenses, number of attackers, attack timeline, discovery rates, and degrees of statistical dependence all interrelate and lead to tradeoffs in designing a defensive posture. Understanding this interplay and these tradeoffs can guide strategic choices and selection of specific defenses to manage risk in a way that is tailored to the organization and its threats. 

Aside from the unsurprising result that playing “whack-a-mole” against many adversaries is ill-advised, defenders have reason to be optimistic. The required increases in number of defenses grow more slowly than the number of attackers. Statistical dependence among those attackers appears to have stronger effect than it does among the defenses. Delay strategies are possible where defenders work more slowly than attackers, at least for small or moderate numbers of attackers. And it is possible, at least theoretically, to think through the optimal defensive postures that match risk tolerance and budget for both defensive strategies.

\bibliographystyle{unsrt}  
\bibliography{references}

\end{document}